\documentstyle[aps,preprint,eqsecnum]{revtex}

\def\endignore{}
\def\ignore #1\endignore{}

\ifx\epsffile\undefined
\def\insertfig#1{}
\else
\def\insertfig#1{{\baselineskip=4pt
\centerline{\epsfxsize=\hsize\epsffile{#1}}}}\fi

\def\eq{\begin{equation}}
\def\eeq{\end{equation}}

\def\eqa{\begin{eqnarray}}
\def\eeqa{\end{eqnarray}}

\def\jref#1#2#3#4{#1 {\bf #2}, #3 (#4)}

\def\NPB#1#2#3{\jref{Nucl.\ Phys.}{B#1}{#2}{#3}}

\def\PLB#1#2#3{\jref{Phys.\ Lett.}{#1B}{#2}{#3}}
\def\PR#1#2#3{\jref{Phys.\ Rep.}{#1}{#2}{#3}}
\def\PRD#1#2#3{\jref{Phys.\ Rev.}{D#1}{#2}{#3}}

\def\PRL#1#2#3{\jref{Phys.\ Rev.\ Lett.}{#1}{#2}{#3}}

\def\PRV#1#2#3{\jref{Phys.\ Rev.}{#1}{#2}{#3}}

\def\to{\mathop{\rightarrow}}

\def\Dsl{\hbox{\kern.1em/\kern-.7000em$D$}} 

\def\scr#1{{\cal #1}}

\def\mybar#1{\kern 0.8pt\overline{\kern -0.8pt#1\kern -0.8pt}\kern 0.8pt}
\def\sla#1{\raise.15ex\hbox{$/$}\kern-.57em #1}
\def\Sla#1{\kern.15em\raise.15ex\hbox{$/$}\kern-.72em #1}

\def\roughly#1{\mathrel{\raise.3ex\hbox{$#1$\kern-.75em%
    \lower1ex\hbox{$\sim$}}}}
\def\lsim{\roughly<}

\def\tr{\mathop{\rm tr}}

\def\bra#1{\langle #1 |}
\def\ket#1{| #1 \rangle}

\def\be{\beta}
\def\ga{\gamma}
\def\De{\Delta}

\def\de{\delta}

\def\La{\Lambda}

\def\om{\omega}

\def\ChPT{\raise.45ex\hbox{$\chi$}PT}

\def\rhs{right-hand side}

\def\MeV{{\rm \ MeV}}
\def\GeV{{\rm \ GeV}}

\begin{document}
\tighten
\preprint{\vbox{
\hbox{HUTP-95/A008, MIT-CTP-2418}
\hbox{hep-ph/9502398}}}
\title{Constraints on light quark masses from the heavy meson spectrum}

\author{Markus A. Luty\medskip}

\address{Center for Theoretical Physics \\
Massachusetts Institute of Technology\\
Cambridge, MA 02139}

\author{Raman Sundrum\medskip}

\address{Lyman Laboratory of Physics \\
Harvard University \\
Cambridge, MA 02138\medskip}

\date{February 1995}

\maketitle

\begin{abstract}
We use the observed $SU(3)$ breaking in the mass spectrum of mesons containing
a single heavy quark to place restrictions on the light quark current masses.
A crucial ingredient in this analysis is our recent first-principles
calculation of the electromagnetic contribution to the isospin-violating
mass splittings.
We also pay special attention to the role of higher-order corrections in
chiral perturbation theory.
We find that large corrections are necessary for the heavy meson data to be
consistent with $m_u =0$.We use the observed $SU(3)$ breaking in the mass
spectrum of mesons containing
a single heavy quark to place restrictions on the light quark current masses.
A crucial ingredient in this analysis is our recent first-principles
calculation of the electromagnetic contribution to the isospin-violating
mass splittings.
We also pay special attention to the role of higher-order corrections in
chiral perturbation theory.
We find that large corrections are necessary for the heavy meson data to be
consistent with $m_u =0$.
\end{abstract}
\pacs{?}

\def\emd{electromagnetic mass difference}
\def\pid{$\pi^+$--$\pi^0$ mass difference}

\section{INTRODUCTION}
The light quark current masses $m_u$, $m_d$, and $m_s$ are among the
fundamental parameters in the standard model of particle interactions, but
an accurate and reliable determination of these parameters remains elusive.
The reason is that these masses are small compared to the mass scale
associated with confinement,
 $\La \sim 1 \GeV$, so that the light quarks are
tightly bound inside hadrons and their mass cannot be measured directly.
However, we can expand around the chiral limit $m_{u,d,s} \to 0$ to obtain
information about current mass {\it ratios} using chiral perturbation theory
\cite{cpt,low,gl}.
The idea is that in the chiral limit, there is a $SU(3)_L \times SU(3)_R$
chiral symmetry that is spontaneously broken to $SU(3)_{L + R}$ at the
scale $\La$.
In this limit, the theory contains 8 Nambu--Goldstone bosons, which are
identified with the light pseudoscalar meson octet ($\pi$, $K$, $\eta$).
The low-energy interactions of these states can be parameterized by an
effective lagrangian with a few unknown parameters \cite{leff}.
In addition, the chiral symmetry is explicitly broken by the light quark
masses and by electromagnetism.
This breaking can be treated perturbatively in the quark masses and the
electromagnetic coupling, and selection rules associated with the chiral
symmetry again tightly constrain the form of these perturbations.

The main difficulty in this approach is that $m_s / \La \sim 20$--$30\%$,
so that higher-order effects can significantly change low-order results.
This is the source of the interesting (and controversial) issue of whether
higher-order corrections can be large enough to allow $m_u = 0$
\cite{kapman,leut,donwyl},
thus solving the the strong $CP$ problem.
This will be a large part of the focus of this paper.

In this paper, we apply the methods of heavy quark symmetry
\cite{heavy,wisgur,latt}
and heavy hadron chiral perturbation theory \cite{gref} to the determination
of the light quark masses from the mass spectrum of mesons containing a single
heavy quark.
These states have quantum numbers $P_q \sim Q\bar{q}$, where $Q = b,c$ and
$q = u,d,s$.
If we could ignore electromagnetic effects and higher-order corrections in the
quark masses, we would obtain
\eq
\label{naive}
R \equiv \frac{m_s - \hat{m}}{m_d - m_u}
\buildrel{\displaystyle ?}\over{=}
\frac{P_s - \hat{P}}{P_d - P_u}
\simeq \cases{20 & for $P = D$ \cr
280 & for $P = B$ \cr}
\eeq
where $\hat{m} \equiv \frac 12 (m_u + m_d)$, {\em etc\/}.
(We use the names of the heavy-meson states to denote their masses.)
The enormous discrepancies between these numbers clearly cannot be accounted
for by the higher-order quark mass corrections.
The reason for these discrepancies is simply that the electromagnetic
contribution to $P_d - P_u$ is numerically comparable to the contribution from
the light quark masses.
This underlines the importance of determining the electromagnetic contribution
to the mass differences.

We will carry out an improved analysis of $R$ making use of the heavy-meson
\emd s computed in ref.~\cite{us}.
(For earlier related work, see ref.~\cite{goityem}.)
Our analysis will also include important non-analytic corrections in the quark
masses \cite{goity,jenkins} and will pay careful attention to the role of
higher-order corrections in chiral perturbation theory.
We will briefly describe the elements of the calculation of the \emd s in the
next section.
The following section contains our analysis of $R$, and the final section
gives our conclusions.

\section{COMPUTATION OF ELECTROMAGNETIC MASS DIFFERENCES}
In ref.~\cite{us}, we computed the electromagnetic mass differences of heavy
mesons in terms of measurable data using techniques similar to those used
in the classic calculation of the \pid\ \cite{pid}.
The basic idea in both calculations is to use dispersion-theoretic
arguments together with the ultraviolet properties of QCD
to relate the \emd\ to measured strong-interaction matrix elements.
We review the main features of our calculation here in order to make this
paper more self-contained.

We begin by writing
\eq
(P_d - P_u)^{\rm (EM)} =
\frac{ie^2}{2} \int\! \frac{d^4 k}{(2\pi)^4} \frac{\De T(p, k)}{k^2 + i0+}
\eeq
where $\De T \equiv T_d - T_u$ is a difference of the (traced) Compton
amplitudes
\eq
T_q(p, k) \equiv i \int\! d^4x\, e^{ik\cdot x}
\bra{P_q(p)} J^\mu(0) J_\mu(x) \ket{P_q(p)},
\eeq
where $J^\mu$ is the electromagnetic current.
It can be shown rigorously that this integral converges \cite{collins,us}.

We consider $\De T$ in the large-$N$ limit, where $N$ is the number of QCD
colors.
In this limit, we can express $\De T$ as a sum of tree graphs in a
theory of (infinitely many) mesons with interactions that are at most
polynomial in momenta \cite{witten}.
(This can be viewed as a rigorous version of the ``resonance dominance''
assumption made in refs.~\cite{pid}.)
The vertices of the meson graphs that determine $\De T$ are directly related
to physical processes.
For example, the graphs in fig.~1b determine the heavy-meson form factors,
while the last two graphs in fig.~1a are related to meson scattering
amplitudes.
Thus, the representation of $\De T$ in the meson theory can be viewed as a
double dispersion relation in $k$ and $p$.

We then impose constraints on the couplings appearing in this sum by demanding
that the matrix elements have the infrared behavior required by electromagnetic
gauge invariance, and the ultraviolet behavior demanded by QCD.
These constraints can be expressed as sum rules relating the couplings
appearing in the sum.
(Two of these sum rules are close analogs of the Weinberg sum rules that
appear in the
calculation of the \pid \cite{wsum}.)

Even after imposing the sum rules, the \emd\ is given by an infinite
(convergent) sum over intermediate states.
We then derive an unsubtracted fixed-$\vec k\,^2$ dispersion relation for
$\De T$ that shows that the contributions of intermediate states with
large mass are suppressed by inverse powers of their mass.
It is therefore sensible to assume that the sum is dominated by the
lowest-lying intermediate states.
That is, we keep the minimum number of intermediate states to saturate
the known asymptotic behavior of the form factors and other matrix elements
that appear.
This turns out to be the $P$ and $P^*$ heavy-meson states (which are
degenerate in the limit $m_Q \to \infty$) and the vector mesons
$\rho$, $\om$, $\rho'$ and $\om'$ that are responsible for ``softening'' the
form factors.
When the sum rules are imposed on this restricted set of intermediate states,
there are sufficiently few free parameters that we are able to obtain numerical
estimates for the \emd s.
The results are
\eq
(B^+ - B^0)^{\rm (EM)} \simeq
\Biggl[ +1.7 - 0.13 \left( \frac{\beta}{1 \GeV^{-1}} \right)
-0.03 \left( \frac{\beta}{1 \GeV^{-1}} \right)^2\, \Biggr] \MeV
+ O(1/m_b^2),
\eeq
\eqa
\frac{1}{4} \left[(D^+ - D^0) + 3(D^{*+} - D^{*0})\right]^{\rm (EM)} &\simeq&
\Biggl[ 2.5 + 0.012 \left( \frac{\be}{1 \GeV^{-1}} \right)^2 \nonumber\\
&&\quad + 0.011 \left( \frac{\be}{1 \GeV^{-1}} \right)^2
\left( \frac{\be - \be'}{0.3 \be} \right) \Biggr] \MeV \\
&&\quad + O(1/m_c^2), \nonumber
\eeqa
\eqa
\left[ (D^{*+} - D^{*0}) - (D^+ - D^0)\right]^{\rm (EM)} &\simeq&
\Biggl[ 0.16 + 0.99 \left( \frac{\be}{1 \GeV^{-1}} \right) \nonumber\\
\label{fine}
&&\quad + 0.015 \left( \frac{\be}{1 \GeV^{-1}} \right)^2
\left( \frac{\be - \be'}{0.3 \be} \right) \Biggr] \MeV \\
&&\quad + O(1/m_c^2). \nonumber
\eeqa
Here $\be$ ($\be'$) measure $P^* P \ga$ ($P^* P^* \ga$) couplings.
Heavy-quark symmetry gives $\be = \be' + O(1/m_Q)$, and experimental
constraints from $D^*$ decays give $0 < -\be \le 5 \GeV^{-1}$ \cite{am,cho}.%
\footnote{Our convention for $\be$ differs by a sign from that of
ref.~\cite{am}.}
Since the isospin splittings in the $P$ and $P^*$ systems are equal in the
heavy-quark limit, we cannot use them to get independent constraints on the
quark masses.
For the $D$ system, we have chosen to work in terms of the
``spin-independent'' combination $D + 3D^*$ \cite{jenkins}, for which the
the largest $\be$-dependent term in the \emd\ cancels.
We will use the $D$ hyperfine splitting $D^* - D$ to give a determination
of $\be$ below.

There are a number of simultaneous approximations made in this computation:
the heavy-quark expansion (including $1/m_Q$ terms), the large-$N$ limit, and
the truncation of intermediate states.
The heavy-quark expansion appears to be under good control in our calculation,
in the sense that the $1/m_Q$ corrections are small compared to the leading
terms.
The large-$N$ limit and truncation of intermediate states are equivalent to
approximations that can be made in the computation of the \pid, which then
works to $30\%$.
We conclude that it is reasonable to expect our estimates of the \emd s to be
accurate to about $30\%$.

We now complete our determination of the \emd s by extracting a value for
$\be$ from the $D$ hyperfine splittings.
To order $m_q / m_Q$ in chiral perturbation theory, one finds \cite{jenkins}
\eqa
(D^{*+} - D^{*0}) - (D^{+} - D^{0}) &=&
 [(D^{*+} - D^{*0}) - (D^{+} - D^{0})]^{\rm (EM)} \nonumber\\
&&\quad + \frac{1}{R}[(D^{*}_s - \hat D^{*}) - (D_s - \hat D)].
\eeqa
Experimentally, $(D^{*}_s - \hat D^{*}) - (D_s - \hat D) = 0.5 \pm 2.5 \MeV$,
and $R \ge 20$.
Thus, the second term on the \rhs\ is negligible, and the isospin hyperfine
splitting is dominated by the electromagnetic contribution.
Using eq.~(\ref{fine}), we obtain
\eq
\label{ourbeta}
\be \simeq -1.6 \GeV^{-1}
\eeq
for reasonable values of $\be'$.
This is consistent with the results of refs.~\cite{am,cho}.
For comparison the constituent quark model prediction is
$\be \simeq - 3 \GeV^{-1}$.
It should be emphasized that our determination has errors $O(1/m_c)$, and
there is a further $O(1/m_c)$ error in using this value for the $B$ system.
Using eq.~(\ref{ourbeta}), we obtain
\eqa
\frac{1}{4} \left[(D^+ - D^0) + 3(D^{*+} - D^{*0})\right]^{\rm (EM)}
&\simeq& 2.5 \MeV, \\
(B^+ - B^0)^{\rm (EM)} &\simeq& 1.8 \MeV.
\eeqa
The \emd s are not very sensitive to $\be$, so despite the uncertainties
discussed above, we expect these values to be correct to about $30\%$.

\section{CONSTRAINTS ON $R$}
We now turn to the analysis of $R$ defined in eq.~(\ref{naive}).
We pay special attention to the role of higher-order corrections in chiral
perturbation theory, since it is known that these can have an important effect
on the determination of the light quark mass ratios \cite{kapman,donwyl}.
Keeping terms of order $m_q$, $m_q^{3/2}$, $m_q/m_Q$, $m_q^2$, and the
corresponding logs,%
\footnote{This corresponds to assuming that $m_s / \La \sim \La / m_Q$.
This probably overestimates the $O(m_q/m_Q)$ corrections for the $B$ system,
but this is harmless because our results are identical if these corrections
are omitted.}
the mass differences of the heavy mesons can be written in the form
\eqa
\label{idiff}
P_d - P_u &=& (P_d - P_u)^{\rm (EM)} + A_0 (m_d - m_u)
+ \De_{d-u}^{(3/2)}
+ \left[ A_1(\mu) + \De_1(\mu) \right] \frac{m_d - m_u}{m_Q} \nonumber\\
&&\quad + \left[ A_2(\mu) + \De_2(\mu) \right] m_s (m_d - m_u)
+ O(m_{u,d}^2 \ln m_s) \\
\label{sdiff}
P_s - \hat{P} &=& A_0 m_s + \De_s^{(3/2)}
+ \left[ A_1(\mu) + \De_1(\mu) \right]
\frac{m_s}{m_Q} \nonumber\\
&&\quad + \left[ A_2(\mu) + \De_2(\mu) \right] m_s^2
+ \left[ A_3(\mu) + \De_3(\mu) \right] m_s^2
+ O(m_s^{5/2}).
\eeqa
(For the $D$ system, $P = \frac{1}{4} (D + 3D^*)$.)
The meaning of these terms is as follows:
$(P_d - P_u)^{\rm (EM)}$ is the electromagnetic contribution discussed in the
previous sections.
$A_0 m_q$ is the term linear in quark masses that gives rise to
eq.~(\ref{naive}).
$\De^{\rm (3/2)}$ are the $O(m_q^{3/2})$ non-analytic corrections, which can be
expressed in terms of the light meson masses and couplings
\cite{goity,jenkins}:%
\footnote{Ref.~\cite{goity} contains an error that was corrected in
ref.~\cite{jenkins}.}
\eqa
\label{nadu}
\De^{(3/2)}_{d-u} &=& -\frac{g^2}{16\pi f^2} (M_{K^0}^3 - M_{K^+}^3
+ \frac{1}{2R} M_\eta^3), \\
\label{nas}
\De^{(3/2)}_s &=&
-\frac{g^2}{16\pi f^2} (M_K^3 + \frac{1}{2} M_\eta^3).
\eeqa
Here, $f \simeq 110 \MeV$ is the $K$ decay constant and $g$ measures the
strength of the $PP\pi$ coupling \cite{gref}.
The experimental limit on the $D^*$ width gives the bound $g^2 \le 0.5$
\cite{gref}.
(To the order we are working, the value of $g$ is the same in the $D$ and $B$
system by heavy quark flavor symmetry.)
However, the analysis of refs.~\cite{am,cho} reveals that the
magnitudes of $g$ and $\beta$ are correlated, so that $g^2 \simeq 0.15$ for
the value of $\beta$ in eq.~(\ref{ourbeta}).
Their analysis also shows that $g^2 \ge 0.1$.
We will conservatively assume $0.1 \le g^2 \le 0.3$, which corresponds to
$-\be \le 3 \GeV^{-1}$.

The terms proportional to $A_1$ and $A_{2,3}$ in eqs.~(\ref{idiff},\ref{sdiff})
arise from analytic counterterms in the effective lagrangian of order
$m_q/m_Q$ and $m_q^2$, respectively.
The terms proportional to $\De_{1,2,3} \sim \ln m_s$ are the chiral logs that
renormalize them.
The counterterms and the chiral logs each depend on a renormalization scale
$\mu$ in such a way that the terms proportional to $A_j + \De_j$ are
independent of $\mu$.
The fact that the chiral logs are proportional to the same function of quark
masses as the corresponding counterterms is a consequence of the simple
structure of the $SU(3)$ group theory for this system.

Perhaps surprisingly, the chiral logs in $\De_{2,3}$ cannot be computed in
terms of known quantities.
The reason is the presence of higher-order terms in the effective heavy-meson
lagrangian such as (in the notation of ref.~\cite{gref})
\eq
\de\scr L = \frac{c}{\La} \tr\left[ H^\dagger H
(\xi^\dagger \partial^\mu \xi - \xi \partial^\mu \xi^\dagger)
(\xi^\dagger \partial_\mu \xi - \xi \partial_\mu \xi^\dagger) \right],
\eeq
that are not constrained by present data.
These terms give rise to non-analytic contributions such as
\eq
\De_3 m_s^2 \sim \frac{c}{\La} \frac{M_K^4}{16\pi^2 f^2}
\ln\frac{M_K^2}{\mu^2}
= O(m_s^2 \ln m_s).
\eeq
(A similar fact was noted for baryons in ref.~\cite{lebedluty}.)
In principle, the log corrections are enhanced over the counterterm
contributions by $\sim \ln M_K^2 / \mu^2$ for $\mu \sim 1 \GeV$, but in
practice the logs are not significantly larger than unity.
We will therefore treat the terms proportional to $A_j + \De_j$ as unknown
corrections.

We can solve for $R$ from eqs.~(\ref{idiff},\ref{sdiff}) to obtain
\eq
R = \frac{(P_s - \hat{P}) - \De^{(3/2)}_s(K) - \bar{A}_3}
{(P_d - P_u) - (P_d - P_u)^{\rm (EM)} - \De^{(3/2)}_{d-u}(K)}.
\eeq
where $\De^{(3/2)}(K)$ are the terms in the $O(m_q^{3/2})$ corrections which
depending on the $K$ masses (see eqs.~(\ref{nadu},\ref{nas}));
$\bar{A}_3 \equiv \left[A_3(\mu) + \De_3(\mu) \right] m_s^2$
parameterizes the unknown $O(m_q^2)$ and $O(m_q^2 \ln m_q)$ deviations
from the $O(m_q^{3/2})$ mass relations for both the $D$ and $B$ systems.
Since such relations are expected to work to $20$--$30\%$,
it is reasonable to assume that $\bar{A}_3$ should not be much larger than
$30\%$.
Ref.~\cite{kapman} advocates a different measure of the chiral corrections.
They demand that the second-order corrections to individual masses be less
than $30\%$, but allow this to be the result of cancellations between
larger corrections.
Following this criterion, we would allow $\bar{A}_3$ to be $60\%$, since this
can cancel against $-30\%$ corrections arising from the term proportional to
$A_2 + \De_2$ to give $30\%$ corrections to $P$ mass differences.

We show $R$ as a function of $(P_d - P_u)^{\rm (EM)}$ in figs.~2 and 3
for values of $\bar{A}_3$ corresponding to $0$, $30\%$, and $60\%$ of
$P_s - \hat{P}$.
We see that if the chiral corrections parameterized by $\bar{A}_3$ are
$\sim 30\%$, and the computed \emd s have errors $\lsim 30\%$, then the
results for both the $D$ and $B$ systems can comfortably accomodate $R = 44$.
This is the value of $R$ predicted by lowest-order chiral perturbation theory
for the light meson masses;
it is also the value obtained by the $O(m_q^2)$ analysis of ref.~\cite{gl}.

We can use our results to address the question of whether $m_u = 0$ by using
a relation between $R$ and $m_u / m_d$ obtained from the light pseudoscalar
masses that is valid to $O(m_q^2)$ \cite{kapman,leut} and that predicts
$R = 24 \pm 2$ for $m_u = 0$.
{}From figs.~2 and 3, we see that if we assume that $\bar{A}_3$ is a $30\%$
correction, we require very large corrections in both the $D$ and $B$ \emd s
in order to be consistent with $m_u = 0$.
If $\bar{A}_3$ is a $60\%$ correction, the data can accomodate $m_u = 0$.

\section{CONCLUSIONS}
We have derived constraints on the light quark mass ratio $R$ from the spectrum
of mesons containing a single heavy quark, using the QCD-based computation of
the heavy-meson \emd s of ref.~\cite{us}.
Our results, summarized in figs.~2 and 3, indicate that even when $30\%$
uncertainties are assigned to both the \emd s and the unknown $O(m_q^2)$
chiral corrections, the value of $R$ is bounded away from the value required
by $m_u = 0$.
While we do not regard this as definitive proof that $m_u \ne 0$, it is
striking that the central values of higher-order analyses of both the
light pseudoscalar mesons and the heavy mesons prefer $m_u \ne 0$, and
large chiral corrections must be invoked in both cases to be consistent with
$m_u = 0$.

\section*{ACKNOWLEDGEMENTS}
We would like to thank J. F. Donoghue, A. Falk,
R. F. Lebed, M. Savage and M. Suzuki for helpful discussions.
We also thank the Institute for Theoretical Physics at Santa Barbara for
hospitality while this work was in progress.
MAL is supported in part by DOE contract DE-AC02-76ER03069 and by NSF
grant PHY89-04035;
RS is supported in part by NSF Grant NSF PHY-92-18167.

\begin{figure}
\label{diagrams}
\caption{Diagrams contributing to $T$ in the large-$N$ limit of QCD.
The sum over $n$ is over heavy mesons with quantum number $Q\bar{q}$, while
the sums over $r$ and $s$ are over light vector mesons with quantum numbers
$q\bar{q}$.}
\end{figure}

\begin{figure}
\label{Dgraph}
\caption{The quark mass ratio $R$ as a function of the spin-independent \emd\
in the $D$ system.
Our prediction for the \emd\ is shown by the vertical dashed line.
The bands correspond to the range $0.1 \le g^2 \le 0.3$.
The chiral corrections parameterized by $\bar{A}_3$ are assumed to be $0\%$
in the upper band, $30\%$ in the middle band, and $60\%$ in the lower band.
The value $R = 44$ is the value obtained from a leading order analysis of
the light meson masses, while the value $R = 24$ is the value required by
$m_u = 0$.}
\end{figure}

\begin{figure}
\label{Bgraph}
\caption{The quark mass ratio $R$ as a function of the $B$ \emd.
Our prediction for the \emd\ is shown by the vertical dashed line.
The bands correspond to the range $0.1 \le g^2 \le 0.3$.
The chiral corrections parameterized by $\bar{A}_3$ are assumed to be $0\%$
in the upper band, $30\%$ in the middle band, and $60\%$ in the lower band.}
\end{figure}


\begin{references}
\bibitem{cpt} M. Gell-Mann, R. Oakes, and B. Renner, \PRV{175}{2195}{1968};
P. Langacker and H. Pagels, \PRD{8}{4595}{1973}.

\bibitem{low} S. Weinberg, in {\it Festschrift for I. I. Rabi}, L. Motz
{\it ed.} (New York Academy of Sciences, New York, 1977).

\bibitem{gl} J. Gasser and H. Leutwyler, \PR{87}{77}{1982};
\NPB{250}{465}{1985}.

\bibitem{leff} J. Schwinger, \PLB{24}{473}{1967};
S. Coleman, J. Wess, and B. Zumino, \PRV{177}{2239}{1969}; C. Callan,
S. Coleman, J. Wess, and B. Zumino, \PRV{177}{2247}{1969};
S. Weinberg,  Physica {\bf 96A}{327}{1979}.

\bibitem{kapman} D. Kaplan and A. V. Manohar, \PRL{56}{2004}{1986}.

\bibitem{leut} H. Leutwyler, \NPB{337}{108}{1990}.

\bibitem{donwyl} J. F. Donoghue, B. R. Holstein, and D. Wyler,
\PRL{69}{3444}{1992};
J. F. Donoghue and D. Wyler, \PRD{45}{892}{1992}.

\bibitem{heavy} S. Nussinov and W. Wentzel, \PRD{39}{130}{1987};
M. B. Voloshin and M. A. Shifman, Sov.\ J.\ Nucl.\ Phys.\ {\bf 45}, 292 (1987);
Sov.\ J.\ Nucl.\ Phys.\ {\bf 47}, 511 (1987).

\bibitem{wisgur} N. Isgur and M. Wise, \PLB{208}{504}{1988};
\PLB{232}{113}{1989}; \PLB{237}{527}{1990}.

\bibitem{latt} W. E. Caswell and G. P. Lepage, \PLB{167}{437}{1986};
G. P. Lepage and B. A. Thacker, Nucl.\ Phys.\ {\bf B4}
(Proc.\ Suppl.), 199 (1988);
E. Eichten and B. Hill, \PLB{234}{511}{1990}.

\bibitem{gref} G. Burdman and J. F. Donoghue, \PLB{280}{287}{1992};
M. B. Wise, \PRD{45}{2188}{1992}; T.M. Yan, H.Y. Cheng, C.Y. Cheung, G.L.
Lin, Y.C. Lin, and H.L. Yu, \PRD{46}{1148}{1992}.

\bibitem{us} M. A. Luty and R. Sundrum, hep-ph/9502259, submitted to
Phys.~Rev.~{\bf D}.

\bibitem{goityem} J. L. Goity and W.-S. Hou, \PLB{282}{243}{1992}.
J. L. Goity, \PLB{303}{337}{1993}.

\bibitem{goity} J. L. Goity, \PRD{46}{3929}{1992}.

\bibitem{jenkins} E. Jenkins, \NPB{412}{181}{1994}.

\bibitem{pid} T. Das, G. S. Guralnik, V. S. Mathur, F. E. Low, and J. E. Young,
\PRL{18}{759}{1967};
for a recent review, see S. Pokorski, {\it Gauge Field Theories}
(Cambridge, 1987).

\bibitem{collins} J. C. Collins, \NPB{149}{90}{1979}.

\bibitem{witten} See {\it e.g.} E. Witten, \NPB{160}{57}{1979}.


\bibitem{wsum} S. Weinberg, \PRL{18}{507}{1967};
C. Bernard, A. Duncan, J. Lo Secco, and S. Weinberg, \PRD{12}{792}{1975}.

\bibitem{am} J. F. Amundson, C. G. Boyd, E. Jenkins, M. Luke, A. V. Manohar,
J. L. Rosner, M. J. Savage, and M. B. Wise, \PLB{296}{415}{1992}.

\bibitem{cho} P. Cho and H. Georgi, \PLB{296}{408}{1992}.


\bibitem{lebedluty} R. F. Lebed and M. A. Luty, \PLB{329}{479}{1994}.

\end{references}
\end{document}